\documentclass[aps,prb,twocolumn,superscriptaddress,showpacs, floatfix]{revtex4-1}

\usepackage{graphicx}
\usepackage{calc}
\usepackage{bm}
\usepackage{color}
\usepackage{mathtools}
\usepackage{amsmath}

\begin{document}

\title{Spin-valve effect for spin-polarized surface states in topological semimetals}

\author{A.A.~Avakyants}
\author{V.D.~Esin}
\author{D.Yu.~Kazmin}
\author{N.N.~Orlova}
\author{A.V.~Timonina}
\author{N.N.~Kolesnikov}
\author{E.V.~Deviatov}

\affiliation{Institute of Solid State Physics of the Russian Academy of Sciences, Chernogolovka, Moscow District, 2 Academician Ossipyan str., 142432 Russia}

\date{\today}

\begin{abstract}
We experimentally investigate magnetoresistance  of a single GeTe-Ni junction between the $\alpha$-GeTe topological semimetal and thick nickel film at room and liquid helium temperatures. For the magnetic field parallel to the junction plane, we demonstrate characteristic spin-valve  hysteresis with mirrored differential resistance $dV/dI$ peaks even at room temperature. In contrast, for normal magnetic fields spin-valve effect  appears only at low temperatures. From the magnetic field anisotropy, observation of the similar effect for another topological semimetal Cd$_3$As$_2$, and strictly flat $dV/dI(H)$ magnetoresistance curves for the reference GeTe-Au junction, we connect the observed spin-valve effect with the spin-dependent scattering between the spin textures in the topological surface states and the ferromagnetic nickel electrode. For the topological semimetal $\alpha$-GeTe, room-temperature spin-valve effect allows efficient  spin-to-charge conversion even at ambient conditions.
\end{abstract}

\maketitle

\section{Introduction}

Recent interest to topological semimetals is mostly connected with topologically protected  surface states. In Weyl semimetals (WSM), Fermi arc surface states connect projections of Weyl nodes on the surface Brillouin zone and these surface states inherit the chiral property of the Chern insulator edge states~\cite{armitage}. Spin-orbit interaction lifts the spin degeneracy of the surface states leading to their in-plane spin polarization, with strongly correlated and predominantly antiparallel spin textures in the neighboring Fermi arcs~\cite{Burkovetal2018}.

Spin- and angle- resolved photoemission spectroscopy  data indeed demonstrate  surface Fermi arcs with nearly full spin polarization~\cite{das16,feng2016,lv2015,xu16}.  Moreover,   complex spin textures  appear~\cite{jiang15,rhodes15,wang16} in WTe$_2$ nonmagnetic WSM due to the spin-momentum locking~\cite{Sp-m-lock}.  Surface topological textures (skyrmions) were also visualized in some magnetic semimetals by STM, Lorenz electron microscopy, and magnetic force microscopy~\cite{CrGeTe,FGT_skyrmion1,FGT_skyrmion2}. Recent investigations show  topological protection  of skyrmion structures due to their origin from the spin-polarized topological  surface states~\cite{Araki}. 

On the other hand,  magnetic skyrmions usually appear in the artificial objects like ferromagnetic multilayers, e.g. for Co/Pd~\cite{Co/Pd}, Pt/Co/Ta~\cite{Pt/Co/Ta,Pt/Co/Ta2}, and  Ir/Fe/Co/Pt~\cite{Ir/Fe/Co/Pt} multilayer films. Multilayers demonstrate two independent magnetization processes~\cite{SrRuOheterostr}, inverted hysteresis~\cite{invhyst}, exchange bias~\cite{exchbias} and spin-valve~\cite{spin valve1,spin valve2} effects. In the latter case, the multilayer resistance depends on the  mutual orientation of the layers' magnetizations due to the spin-dependent scattering, so the spin-valve resistance can be affected by external magnetic field or high current density. For example, current-induced excitation of spin waves, or magnons,  was demonstrated as sharp $dV/dI$ differential resistance  peaks~\cite{myers,tsoi1,tsoi2,katine,single,balkashin,balashov}. 

Due to the different spin polarization of the Fermi arc surface states and ferromagnetic bulk, magnetic topological materials also demonstrate spin-valve effect at relatively low  current densities~\cite{timnal,cosns,bite}. At constant magnetic field, bulk magnons~\cite{weyl_magnon} are excited by the spin polarized current within the topological surface states. Also,  similarly to the case of topological insulators~\cite{topinssurf}, one can expect current-induced magnetization dynamics~\cite{current} also for  surface magnetic textures~\cite{Araki,araki1} in WSM. 

It seems to be reasonable to investigate magnetic-field-induced magnetization dynamics for spin valves based on  non-magnetic topological semimetals, which is enabled by strong spin-orbit coupling and spin-momentum locking in the surface states. The spin valves, based on the topological materials, have also the  application potential in energy-efficient spin-to-charge conversion~\cite{sun-chen}. 

 Among nonmagnetic WSM materials, GeTe is of special interest~\cite{GeTespin-to-charge,GeTereview,GeTeour} due to the reported  giant Rashba splitting~\cite{GeTerashba,Morgenstern,GeTerashba1,GeTeour}.   The direct measurement of the Rashba-split surface states of $\alpha$-GeTe(111) has been experimentally realized thanks to K doping~\cite{GeTesurfStates}. It has been shown that the surface states are not the result of band bending and that they are decoupled from the bulk states. The giant Rashba splitting of the surface states of $\alpha$-GeTe is largely arising from the inversion symmetry breaking in the bulk~\cite{GeTesurfStates}. 

Despite extensive research, the spin-to-charge conversion efficiency remains low at room temperature~\cite{sun-chen}. For the topological semimetal~\cite{ortix,triple-point} $\alpha$-GeTe, nonlinear Hall effect has been demonstrated at 300~K~\cite{GeTe2w}, which is the direct experimental manifestation of finite Berry curvature at room temperatures~\cite{sodemann}. 

Here, we experimentally investigate magnetoresistance  of a single GeTe-Ni junction between the $\alpha$-GeTe topological semimetal and thick nickel film at room and liquid helium temperatures. For the magnetic field parallel to the junction plane, we demonstrate characteristic spin-valve  hysteresis with mirrored differential resistance $dV/dI$ peaks even at room temperature. In contrast, for normal magnetic fields spin-valve effect  appears only at low temperatures.

\section{Samples and technique}

\begin{figure}[t]
\includegraphics[width=1\columnwidth]{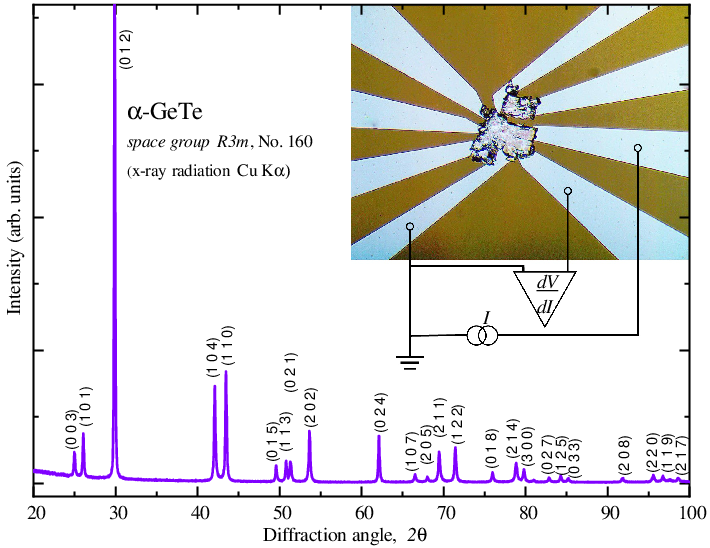}
\caption{(Color online)   X-ray powder diffraction  pattern (Cu K$_{\alpha1}$ radiation), which is obtained for the crushed GeTe single crystal. Single-phase  $\alpha$-GeTe is confirmed with the space group $R3m$ No. 160, so the results below cannot appear from the incorrect stoichiometry or oxides. The  inset shows optical image of the sample: $1~\mu$m thick GeTe flake is placed  on the  $10~\mu$m separated thick Ni ferromagnetic leads. Electrical connections are schematically shown for the  standard three-point technique to investigate magnetoresistance of a single GeTe-Ni junction.
}
\label{fig1}
\end{figure}

We prepare several types of interface structures. The main one are the GeTe-Ni junctions between a ferromagnetic nickel film and a non-magnetic topological semimetal GeTe, see Fig.~\ref{fig1}  (b). For comparison, we also prepare Cd$_3$As$_2$-Ni junctions based on the non-magnetic topological Dirac semimetal Cd$_3$As$_2$ and the reference GeTe-Au junctions with a gold normal layer.

GeTe single crystals were grown by physical vapor transport in the evacuated silica ampule. The initial GeTe load was synthesized by direct reaction of the high-purity (99.9999\%) elements in vacuum. For the  crystals growth, the initial GeTe load serves as a source of vapors: it was melted and kept at 770-780$^\circ$ C for 24 h. Afterward, the source was cooled down to 350$^\circ$ C at the 7.5 deg/h rate. The GeTe crystals grew during this process on the cold  ampule walls  above the source.

Cd$_3$As$_2$ crystals were grown by crystallization of molten drops in the convective counterflow of argon held at 5 MPa pressure. For the source of drops the stalagmometer similar to one described~\cite{growth} was applied. The crystals  sometimes had signs of partial habit of $\alpha$-Cd$_3$As$_2$ tetragonal structure. About one fifth of the drops were  single crystals.

The crystal composition is verified by energy-dispersive X-ray spectroscopy for both materials. The powder X-ray diffraction analysis confirms single-phase GeTe or Cd$_3$As$_2$, see Fig.~\ref{fig1}  as an example. The known structure models~\cite{GeTerashba} are also refined with single crystal X-ray diffraction measurements. 

For both junction types, 10~$\mu$m wide ferromagnetic Ni leads are formed by photolithography and  lift-off technique after thermal evaporation of the 100~nm thick nichel film  on the insulating SiO$_2$ substrate.  The topological semimetals are essentially three-dimensional objects\cite{armitage}, so we have to use relatively thick (above 0.5 $\mu m$) flakes for our samples. Small (about 100~$\mu$m size and 1~$\mu$m thick) Cd$_3$As$_2$ or GeTe flakes can be easily obtained  by the mechanical cleaving method.  A single flake (Cd$_3$As$_2$ or GeTe)    is transferred on top of the metallic leads with $\approx 10\times 10~\mu\mbox{m}^2$ overlap and pressed slightly with another oxidized silicon substrate, no external pressure is needed  afterward, see the inset to Fig.~\ref{fig1}. This procedure provides transparent GeTe-Ni or Cd$_3$As$_2$-Ni junctions, stable in different cooling cycles, which has been also demonstrated before for different materials~\cite{timnal,cosns,cdas,getein}. We also use gold leads pattern for the reference GeTe-Au junctions.

\begin{figure}[t]
\includegraphics[width=\columnwidth]{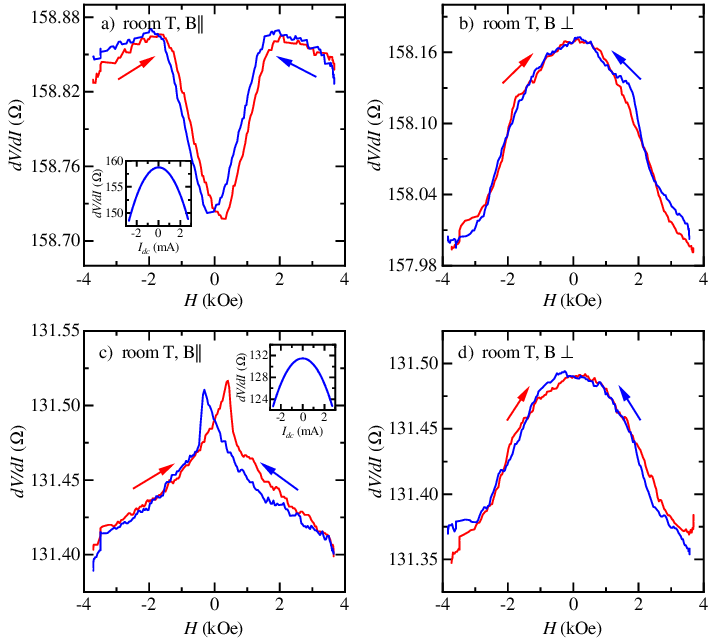}
\caption{(Color online) Magnetoresistance $dV/dI(H)$ of two different GeTe-Ni junctions ((a,b) and (c,d), respectively)  for two opposite magnetic field sweep directions. Insets to (a) and (c) show qualitatively similar  $dV/dI(I)$ current-voltage characteristics for these junctions. The curves are obtained at room temperature, in normal (b,d) and parallel (a,c) to the GeTe-Ni interface magnetic fields.
(b,d) In normal fields, magnetoresistance is always negative, without any  visible hysteresis. 
(a,c) In parallel  magnetic fields,  the sign of the magnetoresistance depends on the particular junction, however, there is well-developed hysteresis for $dV/dI(H)$ curves.   The curves for the opposite sweep directions are centered at $\pm$0.5~kOe, they are mirrored in respect to the zero field, as it can be expected for spin valves. 
  }
\label{fig2}
\end{figure}

We investigate magnetoresistance of a single GeTe-Ni or Cd$_3$As$_2$-Ni junction using a standard three-point technique: one Ni contact is grounded, the neighboring one is used as a voltage probe, while current flows through another contact, as schematically represented in Fig.~\ref{fig1}. To increase the signal/noise ratio, we apply ac current through the sample and  measure ac  voltage  with a lock-in amplifier, also, curves' averaging is necessary. For current-voltage characteristics in the insets to Fig.~\ref{fig2},  an additional dc current $I_{dc}$ is applied, so the ac lock-in signal reflects $dV/dI(I_{dc})$ differential resistance.  For the magnetoresistance measurements,  magnetic field is applied by a standard electromagnet for room temperature  or by superconducting solenoid for liquid helium ($4.2$~K) one.

\begin{figure}
\includegraphics[width=\columnwidth]{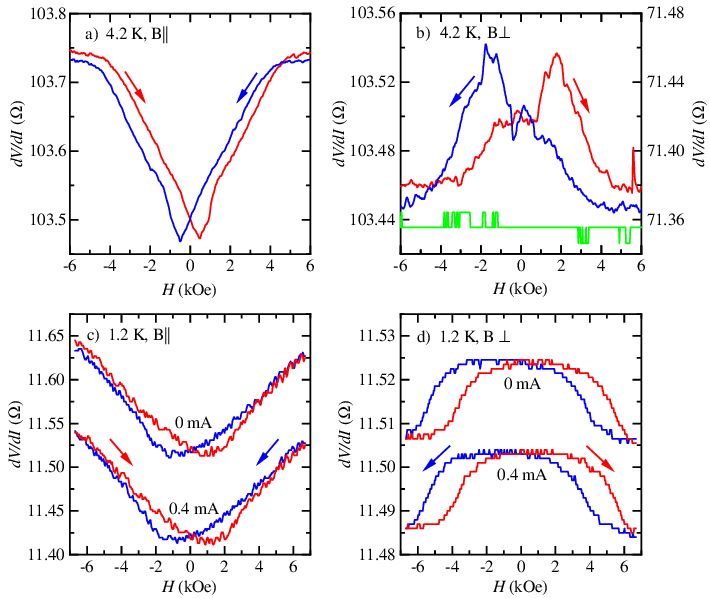}
\caption{(Color online) Magnetoresistance $dV/dI(H)$ of GeTe-Ni and Cd$_3$As$_2$-Ni junctions ((a,b) and (c,d), respectively)  at 4.2~K liquid helium temperature.
(a)  Magnetic field is  parallel to the GeTe-Ni interface, there is no strong temperature dependence for this field orientation, compare with Fig.~\ref{fig2} (a).
(b) In normal magnetic field, two distinct resistance peaks appear also in $\pm 2$~kOe. The $dV/dI(H)$ curves are  mirrored in respect to the zero field, demonstrating spin valve effect. The green  curve is for the reference GeTe-Au junction of approximately similar resistance.
(c,d) Similar spin-valve magnetoresistance for Cd$_3$As$_2$-Ni junction between nickel electrode and the Dirac semimetal Cd$_3$As$_2$. The $dV/dI(H)$ minima and maxima are at $\pm$1.5~kOe for both magnetic field orientations.
}
\label{fig3}
\end{figure} 

\section{Experimental results}

Fig.~\ref{fig2} shows differential magnetoresistance $dV/dI(H)$ of two GeTe-Ni junctions ((a,b) and (c,d), respectively)  for two opposite magnetic field sweep directions. The curves are obtained at room temperature, in normal and parallel to the junction interface magnetic fields. The obtained resistance values are not very different for two samples in Fig.~\ref{fig2} (a) and (c). Moreover, they demonstrate qualitatively similar  $dV/dI(I)$ current-voltage characteristics, see the insets to  Fig.~\ref{fig2} (a) and (c), despite $dV/dI(H)$ curves are even qualitatively different for these junctions in parallel  magnetic field: while magnetoresistance is positive  in  Fig.~\ref{fig2} (a), it is negative in (c). We demonstrate this difference as the maximum device-to-device variation in our experiment.

In a three-point technique, the measured potential $V$ reflects in-series connected resistances of the CeTe-Ni interface,  Ni lead,  and some part of the GeTe crystal flake. The last two contributions are below 1~Ohm for $10^{20}-10^{21}$~cm$^{-3}$ carrier concentration~\cite{GeTe_diamag} in GeTe, which is much below the observed  130-150~Ohm resistance in Fig.~\ref{fig2}. From this fact, and from $dV/dI(I)$ independence of the particular choice of current and voltage probes, we verify that the  interface resistance dominates in the obtained $dV/dI(I)$ and $dV/dI(H)$ curves. Also, $dV/dI(I)$ current-voltage characteristics are typical for semimetal junctions~\cite{orvova_wte2} in the insets to  Fig.~\ref{fig2} (a) and (c). 

For both junctions,  magnetoresistance is always negative if the field is normal to the junction plane, see  Fig.~\ref{fig2} (b) and (d). The curves are qualitatively similar without any visible hysteresis, the overall resistance drop is about 0.15\%. In contrast, there is well-developed hysteresis for $dV/dI(H)$ curves, for opposite sweep directions for these junctions in parallel  magnetic field in  Fig.~\ref{fig2} (a) and (c).  The $dV/dI(H)$ curves are mirrored in respect to the zero field, they are  centered at the same $\pm$0.5~kOe magnetic field for both the junctions with the 0.15\% overall resistance change.  We check, that the observed  $dV/dI(H)$ hysteresis does not depend on the magnetic field sweep rate.  Thus, at room temperature, the experimental $dV/dI(H)$ curves shows typical spin-valve behavior in parallel  magnetic field only, where $dV/dI(H)$ demonstrates a mirror reversal in the opposite field sweeps~\cite{Yao}. 

To our surprise,  the characteristic spin-valve behavior appears for both field orientations at liquid helium temperature, see Fig.~\ref{fig3} (a) and (b). We do not observe strong temperature dependence for the parallel to the interface magnetic field, $dV/dI(H)$ curves are of qualitatively similar behavior  in Figs.~\ref{fig3} (a) and~\ref{fig2} (a) with $dV/dI(H)$ resistance minima at the same $\pm$0.5~kOe.  However, two distinct resistance peaks appear also at $\pm 2$~kOe normal magnetic field, as depicted in Fig.~\ref{fig3} (b). The $dV/dI(H)$ curves are  mirrored in respect to the zero field, the overall resistance drop is still about 0.15\%. We check, that there are no  magnetoresistance or any resistance peaks for the reference GeTe-Au junction of approximately similar resistance, see the green  $dV/dI(H)$ curve in Fig.~\ref{fig3} (b). We should conclude, that the observed spin-valve effect requires thick ferromagnetic Ni layer as an essential part of the structure for the non-magnetic GeTe semimetal. 

The observed behavior is quite universal for the junctions "ferromagnet -- non-magnetic topological semimetal". Very similar spin-valve magnetoresistance can also be demonstrated for Dirac semimetal Cd$_3$As$_2$, see Fig.~\ref{fig3} (c) and (d). In this case, the $dV/dI(H)$ minima and maxima are centered at $\pm$1.5~kOe for both magnetic field orientations.  Thus, despite the qualitative  $dV/dI(H)$ behavior is quite universal for GeTe and Cd$_3$As$_2$ topological semimetals, the exact hysteresis value is defined not only by the Ni electrode, but also by the topological semimetal itself.

\begin{figure}[t]
\includegraphics[width=\columnwidth]{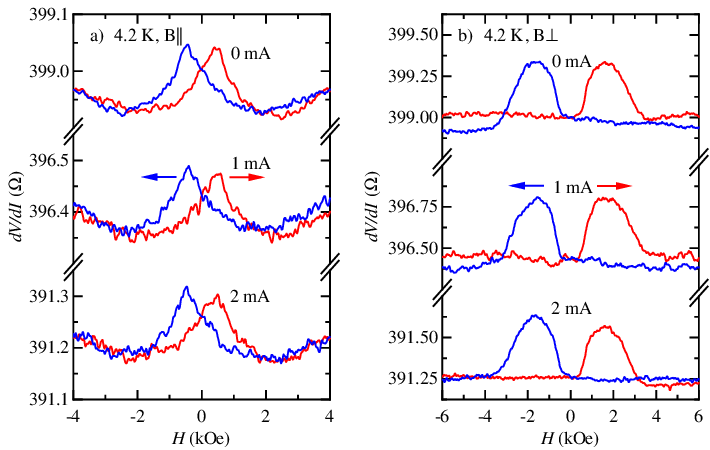}
\caption{(Color online) 
 $dV/dI(H)$ spin-valve hysteresis for the high-resistive GeTe-Ni junction at low 4.2~K temperature. The curves are obtained at different dc currents $I=$0; 1~mA; 2~mA  for two, parallel and normal, orientations of the magnetic field in (a) and (b), respectively. The overall resistance level is diminishing with $I$, while the peaks' amplitudes and their positions are not sensitive to the dc current through the sample.
}
\label{fig4}
\end{figure}

Fig.~\ref{fig4} shows $dV/dI(H)$ hysteresis for another, high-resistive, GeTe-Ni junction at low temperature. The curves are obtained at different dc currents $I$ for two, parallel and normal, orientations of the magnetic field in (a) and (b) respectively. The relative amplitude of the spin-valve effect is smaller for this resistive junction. As it should be expected from current-voltage $dV/dI(I)$ curves, the overall resistance level is diminishing with the dc bias current $I$, while the peaks' amplitudes and their positions are not sensitive to the dc current through the sample.

\section{Discussion} \label{disc}

As a result, we observe pronounced spin-valve effect in  magnetoresistance of the junction between the nonmagnetic topological semimetal and ferromagnetic electrode. For GeTe topological semimetal,  spin-valve hysteresis can even be demonstrated at room temperature  in the parallel to the interface magnetic field.

Spin-valve behavior is usually explained by spin-dependent scattering between two spin-polarized layers. In high magnetic field, spins of two layers are aligned parallel, so the device resistance is at some constant level. While decreasing the magnetic field, parallel alignment and, therefore, the resistance, is preserved until the reversal of the external magnetic field. Afterward, for the opposite field direction, the exact field of the spin reversal depends on the particular layer. Thus, the layers are characterized by mutual  antiparallel  orientation in some magnetic field range, which affects the interlayer spin-dependent scattering, i.e. the device resistance. The sign of the effect depends on the equilibrium mutual orientation of the layers,  and, in general,  on the external field direction. 

Since nickel  is a conventional ferromagnet, the observed spin-valve effect requires some magnetic ordering for the GeTe and Cd$_3$As$_2$  topological semimetals. However, both materials are nominally non-magnetic. For our single crystals, the powder X-ray diffraction analysis confirms pure single-phase GeTe, also, GeTe composition is verified by energy-dispersive X-ray spectroscopy. 
GeTe and Cd$_3$As$_2$  topological semimetals are characterized by diamagnetic bulk, so spin valve effect can not originate from the bulk paramagnetic spin ordering~\cite{Yao}.  The obtained volume susceptibility~\cite{getemag} well corresponds to the known values~\cite{GeTe_diamag}. Thus, any type of magnetic impurities~\cite{magnetic_impurities} can not be responsible for the observed spin-valve effects. Also, we observe qualitatively similar effects for the junctions of very different resistance, so the interface barrier/surface oxides are not responsible for the spin-valve behavior~\cite{Yao}.

There are several characteristic features of $\alpha$-GeTe, which seems to  be important in 
GeTe-Ni devices.  First of all, topological surface states with nontrivial spin textures~\cite{triple-point,GeTesurfStates} can play an important role. In addition, there are nontrivial spin textures with nonzero spin winding numbers in the bulk of $\alpha$-GeTe, which are associated with the type-II Weyl fermions around the triple points of the electronic band structure~\cite{triple-point}. Also, the pronounced spin-orbit splitting~\cite{GeTesurfStates,triple-point} can influence on the charge transport as a whole.

On the other hand, Cd$_3$As$_2$ is not characterized by bulk spin structures, despite Cd$_3$As$_2$-Ni junctions show qualitatively similar spin-valve effects. For this reason, we should consider possible contribution from the surface-state induced spin textures~\cite{GeTesurfStates} in GeTe and Cd$_3$As$_2$  topological semimetals. 

Topological surface states can, in principle, give significant contribution into the spin-dependent transport due to the nearly complete spin polarization of the surface states. For example,  spin polarization of the Fermi arcs can be as high as  $80\%$ in nonmagnetic TaAs~\cite{xu16}. For $\alpha$-GeTe, the giant Rashba splitting of the surface states has been demonstrated, the surface states are decoupled from the bulk~\cite{GeTesurfStates}.

Both the surface-state induced spin textures and 100 nm thick Ni film are characterized by in-plane spin orientation~\cite{Burkovetal2018,GeTesurfStates,getemag}. Thus, both spin-ordered layers can be easily reoriented by low in-plane magnetic fields around $\pm 0.5$~kOe in Figs.~\ref{fig2},~\ref{fig3} and~\ref{fig4}. In this case, the device resistance starts to change before the full reversal of the external magnetic field due to the complicated spin textures within the junction plane~\cite{Burkovetal2018,GeTesurfStates}.  Out-of-plane spin ordering usually requires higher magnetic fields for both spin systems, so reorientation occurs around $\pm$2~kOe in  Figs.~\ref{fig3} and.~\ref{fig4} for normal magnetic fields. 

It is important, that spin-valve effect can not be solely ascribed to the Ni film magnetization reorientation: (i) due to the high Curie temperature, the Ni film is not sensitive to temperature in  4.2~K -- 300~K range, while the spin-valve effect in normal magnetic fields strongly depends on the temperature in this range, compare Figs.~\ref{fig2} and~\ref{fig3}; (ii) the reorientation field range is different for the Cd$_3$As$_2$ Dirac semimetal with helical surface states in Fig.~\ref{fig3}, despite the same Ni film. 

Within the proposed interpretation, one can not expect strong effect of the dc current through the interface in Fig.~\ref{fig4}. While the current could produce bulk spin polarization due to the Edelstein effect~\cite{Edelstein}, the surface states are of nearly complete spin polarization for $\alpha$-GeTe~\cite{GeTesurfStates}.  Thus, the independence of the observed spin-valve effect of the dc current though the structure supports the surface-state origin of the effect.   

For $\alpha$-GeTe, nonlinear Hall effect has been demonstrated at 300~K~\cite{GeTe2w}, which is the direct manifestation of finite Berry curvature at this temperature~\cite{sodemann}. The latter is the origin of the giant Rashba splitting of the surface states~\cite{GeTesurfStates}, leading to the room-temperature spin-valve effect in  Fig.~\ref{fig2} for the in-plane magnetic field. Thus, GeTe-Ni junctions allow room temperature  spin-to-charge conversion mediated by the topological surface states.

\section{Conclusion}
As a conclusion, we experimentally investigate magnetoresistance  of a single GeTe-Ni junction between the $\alpha$-GeTe topological semimetal and thick nickel film at room and liquid helium temperatures. For the magnetic field parallel to the junction plane, we demonstrate characteristic spin-valve  hysteresis with mirrored differential resistance $dV/dI$ peaks even at room temperature. In contrast, for normal magnetic fields spin-valve effect  appears only at low temperatures. From the magnetic field anisotropy, observation of the similar effect for another topological semimetal Cd$_3$As$_2$, and strictly flat $dV/dI(H)$ magnetoresistance curves for the reference GeTe-Au junction, we connect the observed spin-valve effect with the spin-dependent scattering between the spin textures in the topological surface states and the ferromagnetic nickel electrode. For the topological semimetal $\alpha$-GeTe, room-temperature spin-valve effect allows efficient  spin-to-charge conversion even at ambient conditions.

\acknowledgments

We wish to thank S.S~Khasanov for X-ray sample characterization and O.O~Shvetsov for help with the Cd$_3$As$_2$ samples.  

\section{FUNDING}
This work was supported by ongoing institutional funding. No additional grants to carry out or direct this particular research were obtained.

\section{CONFLICT OF INTEREST}
The authors of this work declare that they have no conflicts of interest.

\end{document}